\documentclass[12pt]{umj_eng2}
\usepackage[T1,T2A]{fontenc}
\usepackage[cp1251]{inputenc}
\usepackage[english,russian]{babel}
\usepackage{amsmath}
\usepackage{amssymb}
\usepackage{cite}
\usepackage{amsfonts, srcltx}
\usepackage{lamsarrow}
\usepackage{pb-diagram,pb-lams}

\def\currentyear{0}
\def\currentvolume{0}
\def\currentissue{0}
\firstpage{1}
\renewcommand{\doin}{}

\font\Sets=msbm10

\def\Z {\hbox{\Sets Z}}
\def\res {\hbox{res}\,}

\def\eq#1{\begin{equation}#1\end{equation}}

\def\seqn#1{\begin{align*}#1\end{align*}}
 
\def\seq#1{\begin{equation*}#1\end{equation*}}

\begin{document}

\thispagestyle{empty}

\title[On the integrability of a discrete analogue \dots]
{On the integrability of a discrete analogue \\ of the Kaup–Kupershmidt equation}

\author{R.N. Garifullin, R.I. Yamilov}

\address{Garifullin Rustem Nailevish,
\newline\hphantom{iii} Institute of Mathematics, Ufa Scientific Center, RAS,
\newline\hphantom{iii} Chenryshevsky str. 112,
\newline\hphantom{iii} 450008, Ufa, Russia}

\email{rustem@matem.anrb.ru}

\address{Yamilov Ravil Islamovish,
\newline\hphantom{iii} Institute of Mathematics, Ufa Scientific Center, RAS,
\newline\hphantom{iii} Chenryshevsky str. 112,
\newline\hphantom{iii} 450008, Ufa, Russia}

\email{RvlYamilov@matem.anrb.ru}

\thanks{\rm The research is supported by the Russian Science Foundation (project no. 15-11-20007).}

\maketitle {\small
\begin{quote}
\noindent{\bf Abstract.}
We study a new example of equation obtained as a result of a recent generalized symmetry classification of differential-difference equations defined on five points of one-dimensional lattice. We have established that in the continuous limit this new equation goes into the well-known Kaup–Kupershmidt equation. We have also proved its integrability by constructing an $L-A$ pair and conservation laws. Moreover, we present a possibly new scheme for deriving conservation laws from $L-A$ pairs.

\medskip
\noindent {\bf Keywords:} differential-difference equation, integrability, Lax pair, conservation law

\medskip
\noindent{\bf Mathematics Subject Classification: }{37K10, 35G50, 39A10}

\end{quote}
 }

\section{Introduction}

We consider the differential-difference equation
\eq{u_{n,t}=(u_n^2-1)\left(u_{n+2}\sqrt{u_{n+1}^2-1}-u_{n-2}\sqrt{u_{n-1}^2-1}\right),\label{d_eq}}
where $n\in\Z,$ while $u_n(t)$ is the unknown function of one discrete variable $n$ and one continuous variable $t$, and the index $t$ denotes time derivative. Equation  \eqref{d_eq} is obtained as a result of generalized symmetry classification of five-point differential-difference equations \eq{\label{gen5}u_{n,t}=F(u_{n+2},u_{n+1},u_n,u_{n-1},u_{n-2}),} carried out in \cite{gyl16}. Equation \eqref{d_eq} coincides with the equation \cite[(E17)]{gyl16} up to $u_n$ and $t$ scaling.

Equations \eqref{gen5} play an important role in the study of four-point discrete equations on the square lattice, which are very relevant for today, see e.g. \cite{gy12,a11,mx13,gmy14}. No relation is known between \eqref{d_eq} and any other known equation of the form \eqref{gen5}. More precisely, we mean relations in the form of the transformations \eq{\label{transf}\hat u_n=\varphi(u_{n+k},u_{n+k-1},\ldots,u_{n+m}),\ \ k>m,} and their compositions, see a detailed discussion of such transformations in \cite{gyl16_1}. The only information we have at the moment on \eqref{d_eq} is that it possesses a nine-point generalized symmetry of the form:\seq{u_{n,\theta}=G(u_{n+4},u_{n+3},\ldots,u_{n-4}).}

In this article we explore equation \eqref{d_eq} in details. We have found in Section \ref{s1} its continuous limit, which is the well-known Kaup–Kupershmidt equation \cite{fg80,k80}:
\eq{U_\tau=U_{xxxxx}+5UU_{xxx}+\frac{25}2U_xU_{xx}+5U^2U_x,\label{kk}} where the indices $\tau$ and $x$ denote $\tau$ and $x$ partial derivatives. In order to justify the integrability of \eqref{d_eq}, we construct an $L-A$ pair in Section \ref{s2} and show that it provides an infinity hierarchy of conservation laws in Section \ref{s3}.
In Section \ref{s4} we discuss possible generalizations of a construction scheme for the conservation laws, which  has been formulated in Section \ref{s3} by example of equation \eqref{d_eq}.

\section{Continuous limit}\label{s1}

In a list of equations of the form \eqref{gen5}, presented in \cite{gyl16}, most of equations go in continuous limit into the Korteweg-de Vries equation. The exceptions are \eqref{d_eq} and the following two equations:
\eq{ u_{n,t} = u_n^2(u_{n+2}u_{n+1}-u_{n-1}u_{n-2})-u_n(u_{n+1}-u_{n-1}),\label{seva}}
\eq{u_{n,t} = (u_n+1)\left(\frac{u_{n+2}u_n(u_{n+1}+1)^2}{u_{n+1}}-\frac{u_{n-2}u_n(u_{n-1}+1)^2}{u_{n-1}}+(1+2u_n)(u_{n+1}-u_{n-1})\right)\label{rat},} which correspond to equations (E15) and (E16) of \cite{gyl16}. Equation \eqref{seva} has been known for a long time \cite{th96}. Equation \eqref{rat} has been found recently in \cite{a16_1} and it is related to \eqref{seva} by a composition of transformations of the form \eqref{transf}. These three equations in the continuous limit correspond to the fifth order equations of the form:
\eq{U_{\tau}=U_{xxxxx}+F(U_{xxxx},U_{xxx},U_{xx},U_{x},U).\label{Utau}}

There is a complete list of integrable equations of the form \eqref{Utau}, see \cite{ms12,dss85,mss91}. Two equations play the main role there, namely, \eqref{kk} and the Sawada-Kotera equation \cite{sk74}:
\eq{U_\tau=U_{xxxxx}+5UU_{xxx}+5U_xU_{xx}+5U^2U_x\label{sk}.} All the other are transformed into these two by transformations of the form:
\seq{\hat U=\Phi(U,U_x,U_{xx},\ldots,U_{x\ldots x}).}

It has been known \cite{a11} that equation \eqref{seva} goes in the continuous limit into the Sawada-Kotera equation \eqref{sk}.
The other results below are new. 
Using the substitution
\eq{\label{d_kk}u_n(t)=\frac{2\sqrt 2}3+\frac{\sqrt 2}{16}\varepsilon^2 U\left(\tau-\frac{9}{80}\varepsilon^5 t,x+\frac 23\varepsilon t\right),\quad x=\varepsilon n,} in equation \eqref{d_eq}, we get at $\varepsilon\to 0$ the Kaup–Kupershmidt equation \eqref{kk}.

It is interesting that equation \eqref{rat} has two different continuous limits. 
The substitution 
\eq{\label{r_kk}u_n(t)=-\frac{4}3-\varepsilon^2 U\left(\tau-\frac{18}{5}\varepsilon^5 t,x+\frac 43\varepsilon t\right),\quad x=\varepsilon n,} in \eqref{rat} leads to equation \eqref{kk}, while
the substitution 
\eq{\label{r_sk}u_n(t)=-\frac{2}3+\varepsilon^2 U\left(\tau-\frac{18}{5}\varepsilon^5 t,x+\frac 43\varepsilon t\right),\quad x=\varepsilon n,} leads  to equation \eqref{sk}. As well as \eqref{d_eq}, equation \eqref{rat} deserves further study.

In conclusion, let us present a picture  that shows the link between discrete and continuous equations:
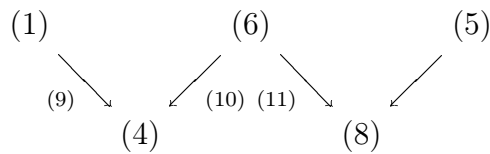
\[ \begin{diagram}
\node{\eqref{d_eq}}
\arrow{se,r}{\eqref{d_kk}} 
\node[2]{\eqref{rat}}
\arrow{sw,r}{\eqref{r_kk}} 
\arrow{se,r}{\eqref{r_sk}} 
\node[2]{\eqref{seva}}
\arrow{sw,r}{} 
 \\
\node[2]{\eqref{kk}}
\node[2]{\eqref{sk}}
\end{diagram}\]

\section{$L-A$ pair }\label{s2}

As the continuous limit shows, equation \eqref{d_eq} should be close to equation \eqref{seva} in its integrability properties. Following the $L-A$ pair \cite[(15,17)]{a11}, we look for an $L-A$ pair of the form:
\eq{L_n\psi_n=0,\quad \psi_{n,t}=A_n\psi_n\label{LA}} with the operator $L_n$ of the form:
\seq{L_n=l_n^{(2)}T^2+l_n^{(1)}T+l_n^{(0)}+l_n^{(-1)}T^{-1},} where $l_n^{(k)},\ k=-1,0,1,2$, depend on the finite number of functions $u_{n+j}.$ Here $T$ is the shift operator:
$Th_n=h_{n+1}.$ In this case the operator $A_n$ can be chosen in the form
\seq{A_n=a_n^{(1)}T+a_n^{(0)}+a_n^{(-1)}T^{-1}.}
The compatibility condition for the system \eqref{LA} has the form:
\eq{\label{L_t}\frac{ d(L_n\psi_n)}{dt}=(L_{n,t}+L_nA_n)\psi_n=0} and it must be satisfied on virtue of equations \eqref{d_eq} and $L_n\psi_n=0$. 

If we suppose that the coefficients $l_n^{(k)}$ depend on $u_n$ only, as in \cite{a11}, then we can see that $a_n^{(k)}$ depend on $u_{n-1},u_n$ only. However, in this case the problem has no solution. Therefore we pass to the case when the functions $l_{n}^{(k)}$ depend on $u_n,u_{n+1}$. Then the coefficients $a_n^{(k)}$ must depend on $u_{n-1},u_n,u_{n+1}$. In this case we have managed to find the operators $L_n$ and $A_n$ with one irremovable arbitrary constant $\lambda$, which plays here the role of spectral parameter:
\eq{L_n=u_n\sqrt{u_{n+1}^2-1}T^2+u_{n+1} T+\lambda\left(u_n-u_{n+1}\sqrt{u_n^2-1}T^{-1}\right),\label{Lsc}}
\eq{A_n=\frac{\sqrt{u_n^2-1}}{u_n}\left(\sqrt{u_n^2-1}(u_{n+1}T+u_{n-1}T^{-1})-\lambda^{-1}u_{n-1}T+\lambda u_{n+1}T^{-1}\right).\label{Asc}}

The $L-A$ pair (\ref{LA},\ref{Lsc},\ref{Asc}) can be rewritten in the standard matrix form with $3\times 3$ matrices $\tilde L_n, \tilde A_n$:
\seq{\Psi_{n+1}=\tilde L_n \Psi_n,\quad \Psi_{n,t}=\tilde A_n \Psi_n.}
Here a new spectral function is given by
\seq{\Psi_n=2^{-n}\left(\begin{array}{c}\frac{\sqrt{u_n^2-1}}{u_n}\psi_{n+1}\\ \psi_n\\ \psi_{n-1}\end{array}\right),}
and the matrices $\tilde{L}_n, \tilde{A}_n$ read:
\eq{\label{Lm}\tilde{L}_n=\left(\begin{array}{ccc}-\frac{1}{\sqrt{u_n^2-1}}&-\frac{\lambda}{u_{n+1}}&\frac{\lambda\sqrt{u_n^2-1}}{u_n}\\ \frac{u_{n}}{\sqrt{u_n^2-1}}&0 &0\\0 &1&0\end{array}\right),}
\eq{\label{Am}\tilde{A}_n=\left(\begin{array}{ccc}
\lambda^{-1}-\frac{u_{n-2}}{u_n}\sqrt{u_{n-1}^2-1}&u_{n+1}\sqrt{u_n^2-1}&\frac{(u_n^2-1)(\lambda u_{n+2}\sqrt{u_{n+1}^2-1}-u_n)}{u_n^2}\\ u_{n+1}\sqrt{u_n^2-1}-\lambda^{-1}u_{n-1}&0 &\frac{\lambda u_{n+1}\sqrt{u_n^2-1}+u_{n-1}(u_n^2-1)}{u_n}\\
u_n+\lambda^{-1}u_{n-2}\sqrt{u_{n-1}^2-1} &u_nu_{n-1}&\lambda+\frac{u_{n-2}}{u_n}\sqrt{u_{n-1}^2-1}\end{array}\right).}
In this case, unlike \eqref{L_t}, the compatibility condition can be represented in matrix form:
\seq{\tilde L_{n,t}=\tilde A_{n+1}\tilde L_n-\tilde L_n \tilde A_n,} without using the spectral function $\Psi_n$.

There are two methods to construct the conservation laws by using such matrix $L-A$ pairs \cite{hy13,gmy14,m15}. However, we do not see how to apply those methods in case of the matrices \eqref{Lm} and \eqref{Am}. In the next section, we will use a different scheme for deriving conservation laws from the $L-A$ pair \eqref{LA}, and that scheme seems to be new.

\section{Conservation laws}\label{s3}

The structure of operators (\ref{Lsc},\ref{Asc}) allows us to rewrite the $L-A$ pair \eqref{LA} in form of the Lax pair. The operator $L_n$ has the linear dependence on $\lambda$:
\eq{L_n=P_n-\lambda Q_n,\label{Lpq}}
where
\seq{P_n=u_n\sqrt{u_{n+1}^2-1}T^2+u_{n+1} T,\quad Q_n=u_{n+1}\sqrt{u_n^2-1}T^{-1}-u_n.}
Introducing $\hat L_n=Q_n^{-1}P_n$, we get an equation of the form:
\eq{\hat L_n \psi_n=\lambda \psi_n.\label{hL}}
The functions $\lambda \psi_n$ and $\lambda^{-1}\psi_n$ in the second equation of \eqref{LA} can be expressed in terms of $\hat L_n$ and $\psi_n$, using \eqref{hL} and its consequence 
$\lambda^{-1} \psi_n=\hat L_n^{-1} \psi_n.$ As a result we have:
\eq{\psi_{n,t}=\hat A_n \psi_n,\label{hA}} where
\seq{\hat A_n=\frac{\sqrt{u_n^2-1}}{u_n}\left(\sqrt{u_n^2-1}(u_{n+1}T+u_{n-1}T^{-1})-u_{n-1}TP_n^{-1}Q_n+ u_{n+1}T^{-1}Q_n^{-1}P_n\right).}

It is important that new operators $\hat L_n$ and $\hat A_n$ in the $L-A$ pair (\ref{hL},\ref{hA}) do not depend on the spectral parameter $\lambda$. For this reason, the compatibility condition can be written in the operator form, without using $\psi$-function:
\eq{\hat L_{n,t}=\hat A_{n}\hat L_n-\hat L_n \hat A_n=[\hat A_n,\hat L_n],\label{comp_h}} i.e. it has now the form of the Lax equation.
The difference between this $L-A$ pair and well-known Lax pairs for the Toda and Volterra equations is that now the operators $\hat L_n$ and $\hat A_n$ are nonlocal. Nevertheless, using the definition of inverse operators, which are linear: \eq{P_n P_n^{-1}=P_n^{-1}P_n=1, \quad Q_nQ_n^{-1}=Q_n^{-1}Q_n=1,\label{invop}}  we can check that \eqref{comp_h} is true by direct calculation.

The conservation laws of equation \eqref{d_eq}, which are expressions of the form 
\seq{\rho_{n,t}^{(k)}=(T-1)\sigma_n^{(k)},\ k\geq 0,}
can be derived from the Lax equation \eqref{comp_h}, notwithstanding nonlocal structure of the operators $\hat L_n,\hat A_n$, see \cite{y06}. For this we must, first of all, represent the operators $\hat L_n,\ \hat A_n$ as formal series in powers of $T^{-1}$:
\eq{H_n=\sum_{k\leq N}h_n^{(k)}T^k.\label{opH}} 
Formal series of this kind can be multiplied according the rule: $(a_nT^k)(b_nT^j)=a_nb_{n+k}T^{k+j}.$ The inverse series can be obtained by definition \eqref{invop}, for instance: 
\seq{Q_n^{-1}=-(1+q_nT^{-1}+(q_nT^{-1})^2+\ldots+(q_nT^{-1})^k+\ldots)\frac {1}{u_n},\ \ q_n=\frac{u_{n+1}}{u_n}\sqrt{u_n^2-1}.}
The series $\hat L_n$ has the second order:
\seq{\hat L_n=\sum_{k\leq 2}l_n^{(k)}T^k=-(\sqrt{u_{n+1}^2-1}T^2+u_{n+1}u_n T+u_{n+1}u_{n-1}\sqrt{u_n^2-1}+\ldots).}

The conserved densities $\rho_{n}^{(k)}$ of equation \eqref{d_eq} can be found as: \eq{\rho_n^{(0)}=\log l_n^{(2)},\qquad \rho_n^{(k)}=\res \hat L_{n}^k,\ \  k\geq 1, \label{form_d}} where the residue of  formal series \eqref{opH} is defined by the rule: $\res H_n=h_n^{(0)}$, see \cite{y06}. Corresponding functions $\sigma_n^{(k)}$ can easily be found by direct calculation.

Conserved densities $\hat \rho_n^{(k)}$ below have been found in this way and then simplified in accordance with the rule:
\seq{\hat\rho_n^{(k)}=c_k\rho_n^{(k)}+(T-1)g_n^{(k)}, } where $c_k$ are constant. First three densities of equation \eqref{d_eq} read:
\seq{\hat\rho_n^{(0)}=\log(u_n^2-1),}
\seq{\hat\rho_n^{(1)}=u_{n+1}u_{n-1}\sqrt{u_n^2-1},}
\seqn{\hat\rho_n^{(2)}&=(u_n^2-1)(2u_{n+2}u_{n-2}\sqrt{u_{n+1}^2-1}\sqrt{u_{n-1}^2-1}+u_{n+1}^2u_{n-1}^2)\\&+u_{n+1}u_{n-1}u_n\sqrt{u_n^2-1}(u_{n+2}\sqrt{u_{n+1}^2-1}+u_{n-2}\sqrt{u_{n-1}^2-1})\nonumber.}

\section{Discussion of the construction scheme}\label{s4}
In previous section we have outlined a construction scheme for the conservation laws by example of equation \eqref{d_eq}. It can easily be generalized to equations of an arbitrarily high order:
\seq{\label{genN}u_{n,t}=F(u_{n+M},u_{n+M-1},\ldots,u_{n-M}).} Let such equation have an $L-A$ pair of the form \eqref{LA} with a linear in $\lambda$ operator $L_n$, and let the operators $P_n,Q_n$ of \eqref{Lpq} have the form:
\eq{R_n=\sum_{k=k_1}^{k_2}r_n^{(k)}T^k,\ \  k_1\leq k_2\in\Z,\label{rop}}
with the coefficients $r_n^{(k)}$ depending on the finite number of functions $u_{n+j}.$ We require that 
\seq{\hat L_n=Q_n^{-1}P_n=\sum_{k\leq N}l_{n}^{(k)}T^k} 
has a positive order  $N\geq 1$. If $N\leq -1$, then we change $\lambda\to\lambda^{-1}$ and introduce $\tilde L_n=P_n^{-1}Q_n$ of a positive order. In case $N=0$ the scheme does not work.

As $\lambda^k\psi_n=\hat L_n^k\psi_n$ for any integer $k$, we can consider operators $A_n$ of the form:
\seq{A_n=\sum_{k= m_1}^{m_2} a_n^{(k)}[T]\lambda^k,\ \ m_1\leq m_2\in\Z,} where $a_n^{(k)}[T]$ are operators of the form \eqref{rop}. Then we can rewrite $A_n$ as:
\seq{\hat A_n=\sum_{k= m_1}^{m_2}a_n^{(k)}[T]L_n^k=\sum_{k\leq \hat N}\hat a_n^{(k)}T^k.}
We are led to the Lax equation \eqref{comp_h} with $\hat L_n,\hat A_n$ of the form \eqref{opH} and, therefore, we can construct the conserved densities as written above, namely, in accordance with \eqref{form_d} with the only difference: $\rho_n^{(0)}=\log l_n^{(N)}$.

It should be remarked that the scheme can easily be applied to equation \eqref{seva} with the $L-A$ pair \cite[(15,17)]{a11}.

This scheme can also be applied in a quite similar way in the  continuous case, namely, to PDEs of the form
\seq{u_t=F(u,u_x,u_{xx},\ldots,u_{x\ldots x}).} We consider the operators \eqref{rop} with $D_x$ instead of $T$, which become the differential operators, where $D_x$ is the operator of total $x$-derivative. Besides, $k_2\geq k_1 \geq 0$ and the coefficients $r_n^{(k)}$ depend on a finite number of functions $u,u_x,u_{xx},\ldots.$ Instead of \eqref{opH} we consider the formal series in powers of $D_x^{-1}.$ A theory of such formal series and, in particular, a definition of the residue are discussed in \cite{msy87}. 

\selectlanguage{english}

\bigskip \renewcommand\Refname{References}

\selectlanguage{russian}


\begin{thebibliography}{99}

\bibitem{a11} V.E. Adler. {\it On a discrete analog of the Tzitzeica equation} // arXiv:1103.5139 [nlin.SI].
 
\bibitem{a16_1}V.E. Adler, \emph{Integrable M\"obius invariant evolutionary lattices of second order}, arXiv:1605.00018 [nlin.SI]

\bibitem{dss85} V.G. Drinfel’d, S.I. Svinolupov, V.V. Sokolov. {\it Classification of fifth order evolution equations
possessing infinite series of conservation laws} // Dokl. AN USSR {\bf  A10.}  7--10 (1985) (in Ukrainian).


\bibitem{fg80} A.P. Fordy and J. Gibbons. {\it Factorization of operators I. Miura transformations} // J. Math. Phys. {\bf 21}.  2508--2510 (1980). http://dx.doi.org/10.1063/1.524357

\bibitem{gmy14} R.N. Garifullin, A.V. Mikhailov and R.I. Yamilov. {\it Discrete equation on a square lattice with a nonstandard structure of generalized symmetries} // Teoret. Mat. Fiz. {\bf 180}:1. 17--34 (2014) [Engl. trans.: Theor. Math. Phys. {\bf 180}:1. 765--780 (2014)]. 

\bibitem{gy12}	R.N. Garifullin and R.I. Yamilov. {\it Generalized symmetry classification of discrete equations of a class depending on twelve parameters} // J. Phys. A: Math. Theor. {\bf 45} (2012) 345205 (23pp). 

\bibitem{gyl16_1} R.N. Garifullin, R.I. Yamilov and D. Levi. {\it Non-invertible transformations of differential-difference equations} //  J. Phys. A: Math. Theor. {\bf 49} (2016) 37LT01 (12pp). 

\bibitem{gyl16}  R.N. Garifullin, R.I. Yamilov, D. Levi. {\it Classification of five-point differential-difference equations} // arXiv:1610.07342 [nlin.SI].

\bibitem{hy13}I.T. Habibullin, M.V. Yangubaeva. {\it Formal diagonalization of a discrete Lax operator and conservation laws and symmetries of dynamical systems} //  Teoret. Mat. Fiz. {\bf 177}:3. 441--467 (2013)
[Engl. trans.: Theor. Math. Phys. {\bf 177}:3. 1655--1679 (2013).]

\bibitem{k80} D.J. Kaup. {\it On the Inverse Scattering Problem for Cubic Eigenvalue Problems of the Class $\psi_{xxx}+6Q\psi_x+6R\psi=\lambda\psi$} // Stud. Appl. Math. {\bf 62}. 189--216 (1980).

\bibitem{ms12} A.G. Meshkov, V.V. Sokolov. {\it Integrable evolution equations with a constant separant} // Ufimsk. Mat. Zh. {\bf 4}:3. 104--154 (2012) [Engl. trans.: Ufa Math. Journal {\bf 4}:3. 104--152 (2012).]

\bibitem{m15} A.V. Mikhailov. {\it Formal diagonalisation of Lax–Darboux schemes} // Model. Anal. Inform. Sist.  {\bf 22}:6. 795–817 (2015), arXiv:1512.07664.

\bibitem{msy87}A.V. Mikhailov, A.B. Shabat and R.I. Yamilov. {\it The symmetry
approach to the classification of nonlinear equations. Complete lists of
integrable systems} // Uspekhi Mat. Nauk. {\bf 42}:4.  3--53 (1987)
[Engl trans.: Russian Math. Surveys {\bf 42}:4. 1--63 (1987).]


\bibitem{mss91}A.V. Mikhailov, V.V. Sokolov and A.B. Shabat. {\it The symmetry approach to classification of integrable
equations} // What is Integrability? Ed. V.E. Zakharov. Springer series in Nonlinear Dynamics. 1991. P.115--184.

\bibitem{mx13}A.V. Mikhailov  and P. Xenitidis. {\it Second order integrability conditions for difference equations: an integrable equation} // Letters in Mathematical Physics {\bf 104}:4. 431--450 (2014) doi:10.1007/s11005-013-0668-8. 

\bibitem{sk74}K. Sawada and  T. Kotera. {\it A method for finding {$N$}-soliton solutions of the {K}.d.{V}. equation and {K}.d.{V}.-like equation} // Progr. Theoret. Phys. {\bf 51.} 1355--1367 (1974).

\bibitem{th96} S. Tsujimoto and R. Hirota. {\it Pfaffian Representation of Solutions to the Discrete BKP Hierarchy in Bilinear Form} // J. Phys. Soc. Jpn. {\bf 65.} 2797--2806 (1996).

\bibitem{y06}
R. Yamilov. {\it{Symmetries as integrability criteria for differential difference equations}} // J. Phys. A: Math. Gen. {\bf 39.} R541--R623 (2006). 

\end{thebibliography}
\end{document}